\documentclass[a4paper]{jpconf}
\usepackage{graphicx}

\usepackage{amsmath,amssymb,yllmath2}
\usepackage[usenames,dvipsnames]{color}
\definecolor{gray}{gray}{0.5}
\providecommand{\omegados}{{\omega_\text{dos}}}

\begin{document}
\title{Emergent granularity and pseudogap near the superconductor-insulator transition}

\author{Nandini Trivedi$^1$, Yen Lee Loh$^2$, Karim Bouadim$^3$, and Mohit Randeria$^1$}

\address{$^1$Department of Physics, The Ohio State University, Columbus, OH  43210, USA}
\address{$^2$Department of Physics and Astrophysics, University of North Dakota, Grand Forks, ND  58202, USA}
\address{$^3$Institute for Theoretical Physics III, University of Stuttgart, Germany}

\ead{trivedi.15@osu.edu}

\begin{abstract}
In two dimensions there is a direct superconductor-to-insulator quantum phase transition driven by increasing disorder. We elucidate, using a combination of inhomogeneous mean field theory and quantum Monte Carlo techniques, the nature of the phases and the mechanism of the transition. We make several testable predictions specifically for local spectroscopic probes.
With increasing disorder, the system forms superconducting blobs on the scale of the coherence length embedded in an insulating matrix. In the superconducting state,
the phases on the different blobs are coherent across the system whereas in the insulator long range phase coherence is disrupted by quantum fluctuations.
As a consequence of this emergent granularity, we show that the single-particle energy gap in the density of states survives across the transition, but coherence peaks exist only in the superconductor.  A characteristic pseudogap persists above the critical disorder and critical temperature, in contrast to conventional theories. Surprisingly, the insulator has a two-particle gap scale that vanishes at the SIT, despite a robust single-particle gap.
\end{abstract}

\section{Introduction}
\begin{figure*}[!h]
\centering
	\begin{minipage}{35pc}
	\includegraphics[width=35pc,trim = 0 31pc 0 6pc, clip]{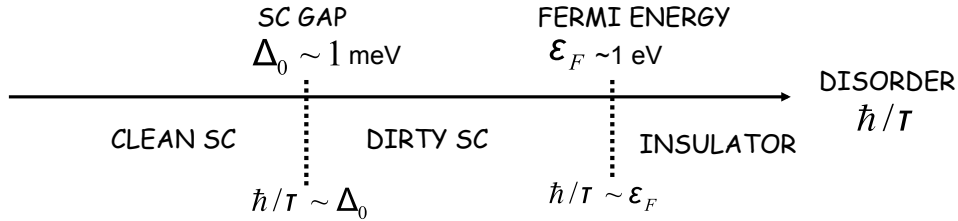}
	\caption{\label{SchematicScales} Typical Energy Scales: A superconductor has a gap of order  $\Delta_0\approx 1\, \mathrm{meV}$, which is much smaller than the Fermi energy $ \epsilon_F\approx 1 \,\mathrm{eV}$. 
	Disorder introduces an energy scale $\hbar/\tau$ where $\tau$ is the mean scattering time. As the disorder is increased $\tau$ gets smaller and the disorder energy scale increases. When $\hbar /\tau\approx \Delta_0$ 
	the system evolves from a clean superconductor to a ``dirty superconductor" but superconductivity remains robust. It is only when $\hbar/\tau$ approaches the Fermi energy that the systems starts to show the superconductor to insulator phase transition.}
	\end{minipage}
\end{figure*}

A superconductor (SC) is an emergent state of matter in which electrons pair up forming Cooper pairs, the different Cooper pairs become phase coherent, and the system undergoes Bose-Einstein condensation. What is the effect of disorder on such a phase-coherent state? It was argued by Anderson~\cite{anderson1959} that three-dimensional superconductivity is quite robust, persisting even in polycrystalline or amorphous materials. Two dimensions turns out to be particularly intriguing because it is the marginal dimension for localization and superconductivity. One can ask: does a two-dimensional system develop superconducting behavior when all its eigenstates are localized due to a random potential and if so, what is the mechanism~\cite{ma1985}?  It is seen from experiments that superconductivity in two dimensions does exist but can be destroyed by a large variety of tuning parameters including temperature, inverse thickness (characterized by sheet resistance), disorder, gate voltage, Coulomb blockade, perpendicular magnetic field, and parallel magnetic field~\cite{strongin1970,haviland1989,valles1989,valles1992,hebard1990,shahar1992,chervenak1999,steiner2005,stewart2007,nguyen2009,lee2011,lin2011,bollinger2011}. Once superconductivity is destroyed the system becomes an insulator beyond a quantum phase transition at $\epsilon_F\tau\approx 1$ (see Fig.~\ref{SchematicScales}).
Much progress has made in the field of superconductor-insulator transitions (SIT) over the years, however, there are still several open questions:

\begin{itemize}
\item For zero disorder, we know that above the superconducting transition temperature $T_c$ the system is a normal Fermi liquid.  What is the nature of the state above $T_c$ at finite disorder?
Is it a Fermi liquid, or a quantum critical phase with a pseudogap?

\item What is the nature of the insulator?  Is it a localized Anderson insulator, a Mott insulator, a Fermi glass, a Bose glass, or something else?

\item What are the energy scale(s) in the insulator that vanish at the SIT?

\item What is the mechanism that drives the SIT?

\item How do the single-particle spectral functions and dynamical conductivity behave in the superconductor, the insulator, and near the SIT?

\item What is the universality class (critical exponents, amplitude ratios, and scaling functions)?

\item Is the critical resistance universal and equal
to the quantum of resistance $R_Q=h/(2e)^2$?  If so (as in Ref.~\cite{bollinger2011}), does this indicate self-duality?

\item Can one develop a theory that captures both Coulomb amplitude suppression and phase fluctuations, and thus unify the fermionic and bosonic pictures of the SIT?

\item Does the SIT happen concomitantly with the metal-insulator transition (MIT), or not?

\item What is the origin of the gigantic peak in the magnetoresistance in the insulating state?
\end{itemize}

In our work so far we have successfully addressed the first five points and we summarize those in this article.

\section{Quantum Phase Transitions}
A SIT is an excellent example of a continuous quantum phase transition (QPT) driven by quantum zero-point fluctuations controlled by a parameter $V$. For  $V < V_c$  the system is a superconductor with well-defined Bogoliubov quasiparticles and collective modes of a superconductor, whereas for $V > V_c$ the system is an insulator, also with well-defined excitations. On the superconducting side, there is an energy scale that vanishes at $V_c$ -- the 2D superfluid density or phase stiffness. Similarly, on the insulating side, one generally expects an energy scale that vanishes at $V_c$.

\begin{figure}[h]
\centering
\begin{minipage}{16pc}
\includegraphics[width=16pc]{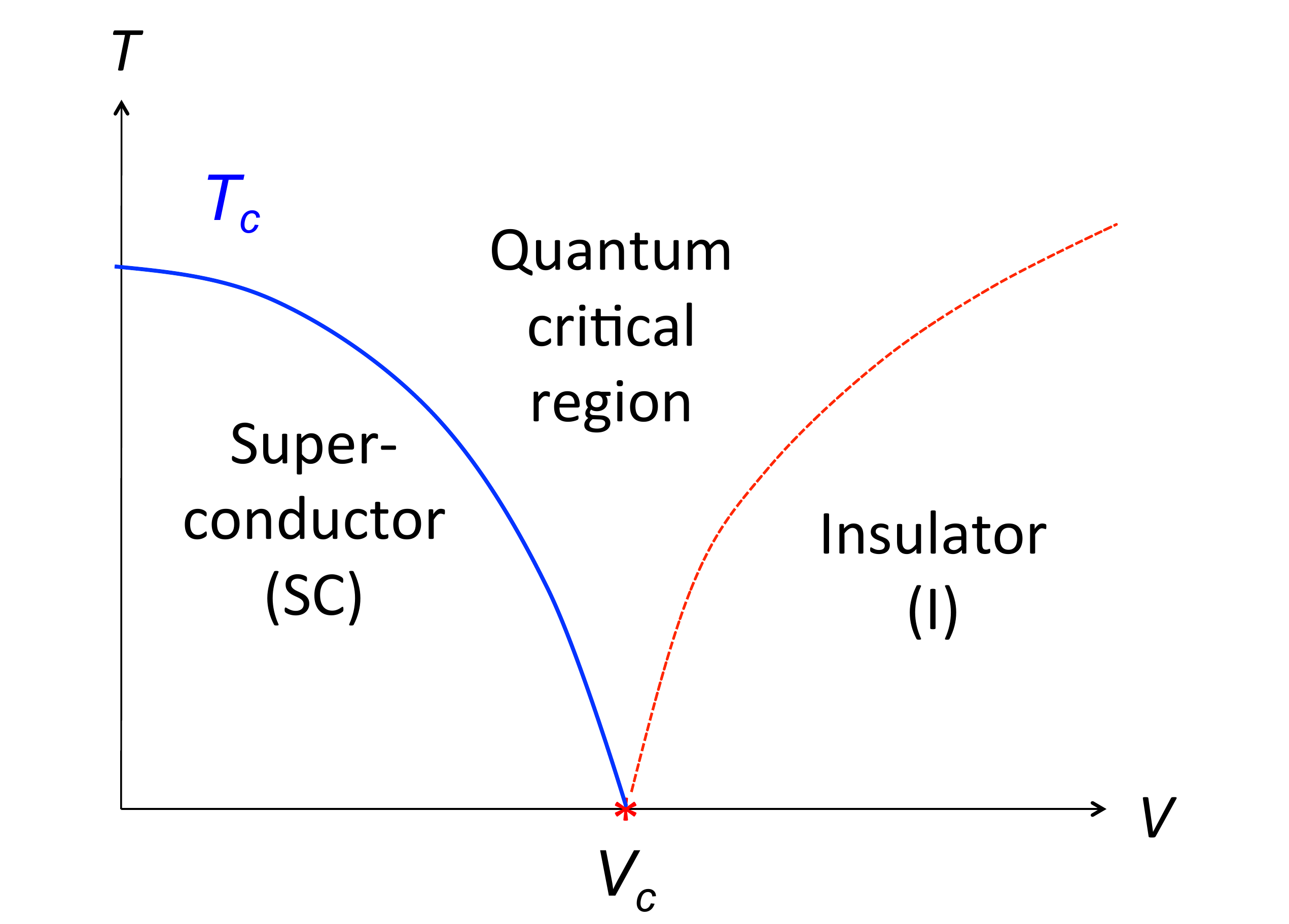}
\caption{\label{qpt} The SIT is a quantum phase transition occurring at zero temperature ($T=0$) and at a critical value of a tuning parameter $V=V_c$.
Above the quantum critical point there is a fan-shaped quantum critical (QC) region where fluctuations are enhanced and QC scaling is obeyed.
}
\end{minipage}
\hspace{2pc}
\begin{minipage}{16pc}
\includegraphics[width=16pc]{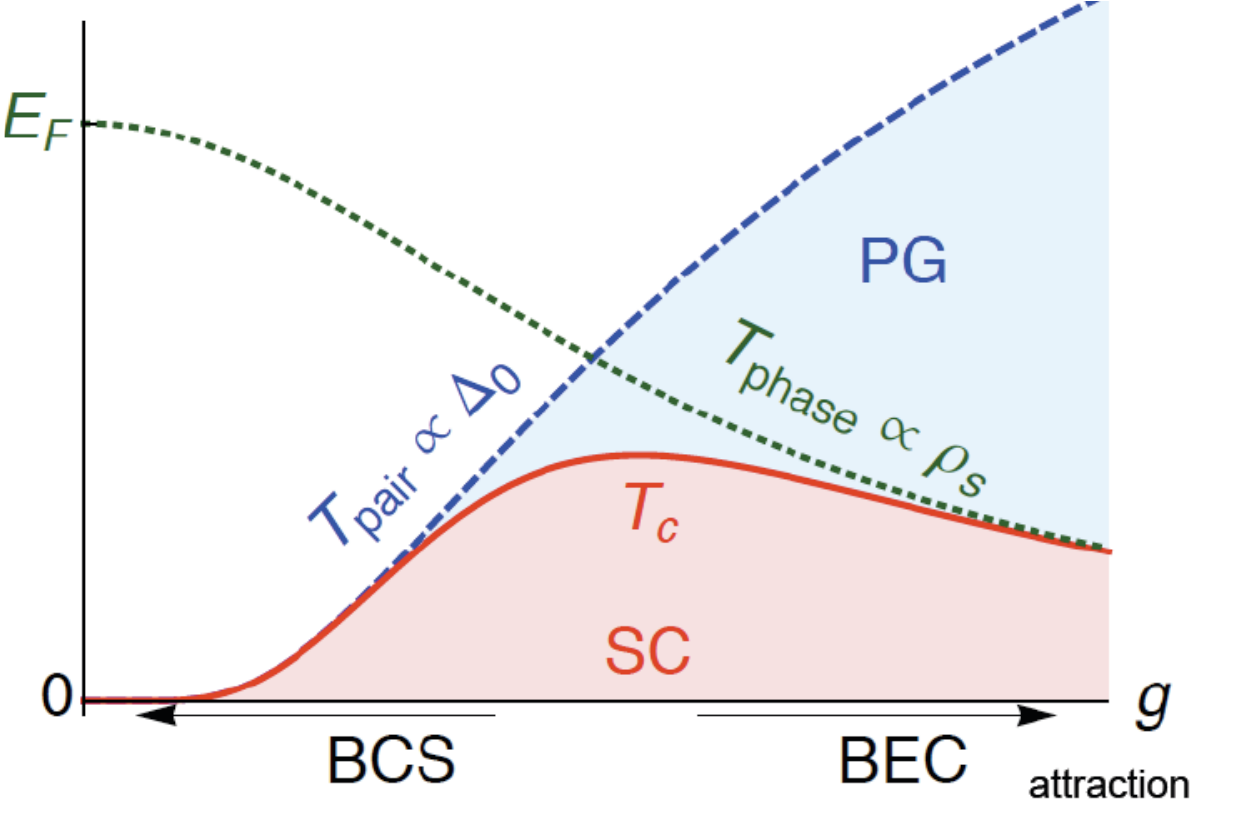}
\caption{\label{bcsbec}BCS-BEC crossover showing the energy and temperature scales of a generic 2D superconductor as a function of
attractive interaction $g$.
}
\end{minipage} 
\end{figure}

The QPT also profoundly affects behavior of the system at finite temperature. The superconducting transition temperature $T_c$ decreases with increasing $V$, 
finally vanishing at the quantum critical point (QCP) at $V = V_c$, as illustrated schematically in Fig.~\ref{qpt}. The fan emanating from the QCP is a quantum critical region that is neither superconducting
nor insulating, where the spectrum is broad and quasiparticles are ill-defined. The primary aim of experimental and theoretical research in this field is to clearly elucidate the properties of the phases and excitations, and then go on to understand the quantum critical region.

\section{Amplitude and Phase Fluctuations}

In order to understand the role of amplitude and phase fluctuations, it is useful to consider a clean problem of fermions with attractive interactions.
A singlet $s$-wave superconductor is described by a complex order parameter 
$\Delta(\RRR) = \left| \Delta(\RRR) \right|  e^{i\theta(\RRR)}$.
At zero temperature the pairing amplitude $\left| \Delta(\RRR) \right|$ takes a uniform value, $\Delta_0$.
This is the energy scale associated with pairing.
It typically manifests itself as an energy gap $E_{g}=\Delta_0$,
and it also sets the maximum temperature, $T_\text{pair} = 0.57 \Delta_0$, 
for the formation of Cooper pairs.
On the other hand, the fluctuations of the phase, $\theta(\RRR)$, are controlled by the superfluid density (or phase stiffness) $\rho_s$.  For 2D superconductors $\rho_s$ has the dimensions of energy, and it can be directly interpreted as the energy scale for phase fluctuations.
$\rho_s$ can be measured using mutual inductance techniques.
It also sets the maximum temperature, $T_\text{phase}$, for long-range phase coherence.

As seen in Fig.~\ref{bcsbec}, with increasing attraction the pairing scale $T_\text{pair} \propto \Delta_0$ increases, whereas the phase coherence scale $T_\text{phase} \propto \rho_s$ decreases (because tightly bound pairs have a larger effective mass in a lattice).
The transition temperature $T_c$ is determined by the lower of the two scales $T_\text{pair}$ and $T_\text{phase}$. 
Weak-coupling superconductors such as Al are described by the BCS (Bardeen-Cooper-Schrieffer) limit of large overlapping Cooper pairs, where $T_\text{pair} \ll T_\text{phase}$.
The phase coherence scale is of the order of the Fermi energy, $T_\text{phase} \sim E_F \sim 10^4 \,\mathrm{K}$, whereas the pairing scale is exponentially suppressed, $T_\text{pair} \sim 1 \,\mathrm{K}$.
Thus, the critical temperature is determined by pairing, and we have a mean-field like, amplitude-driven transition.
Conversely, in the BEC (Bose-Einstein condensation) limit where $T_\text{pair} \gg T_\text{phase}$, the Cooper pairs are small and tightly bound, and the transition is determined by the temperature $T_\text{phase}$ above which phase coherence is lost.
However, a pseudogap -- a suppression in the density of states around the Fermi energy -- persists up to a higher temperature $T_\text{pair}$~\cite{MRNT1992,NTMR1995,randeriaInGreenBook1995,randeriaNewsAndViews2010}.

Although the crossover from BCS to BEC regimes is not directly relevant to superconductivity in weakly coupled traditional superconductors, we will see that as disorder is introduced, the amplitude-driven mechanism for the loss of superconductivity is modified to a phase-driven mechanism opening up a pseudogap regime.
This manifests itself as the discrepancy between transport and tunneling measurements (Fig.~\ref{KarimPhaseDiagram}):
transport measurements are sensitive to global phase coherence and hence to $T_\text{phase}$, whereas tunneling densities of states are sensitive to local pairing correlations and hence to $T_\text{pair}$.

\begin{figure}[h]    \centering
	\begin{minipage}{16pc}
	\includegraphics[width=16pc]{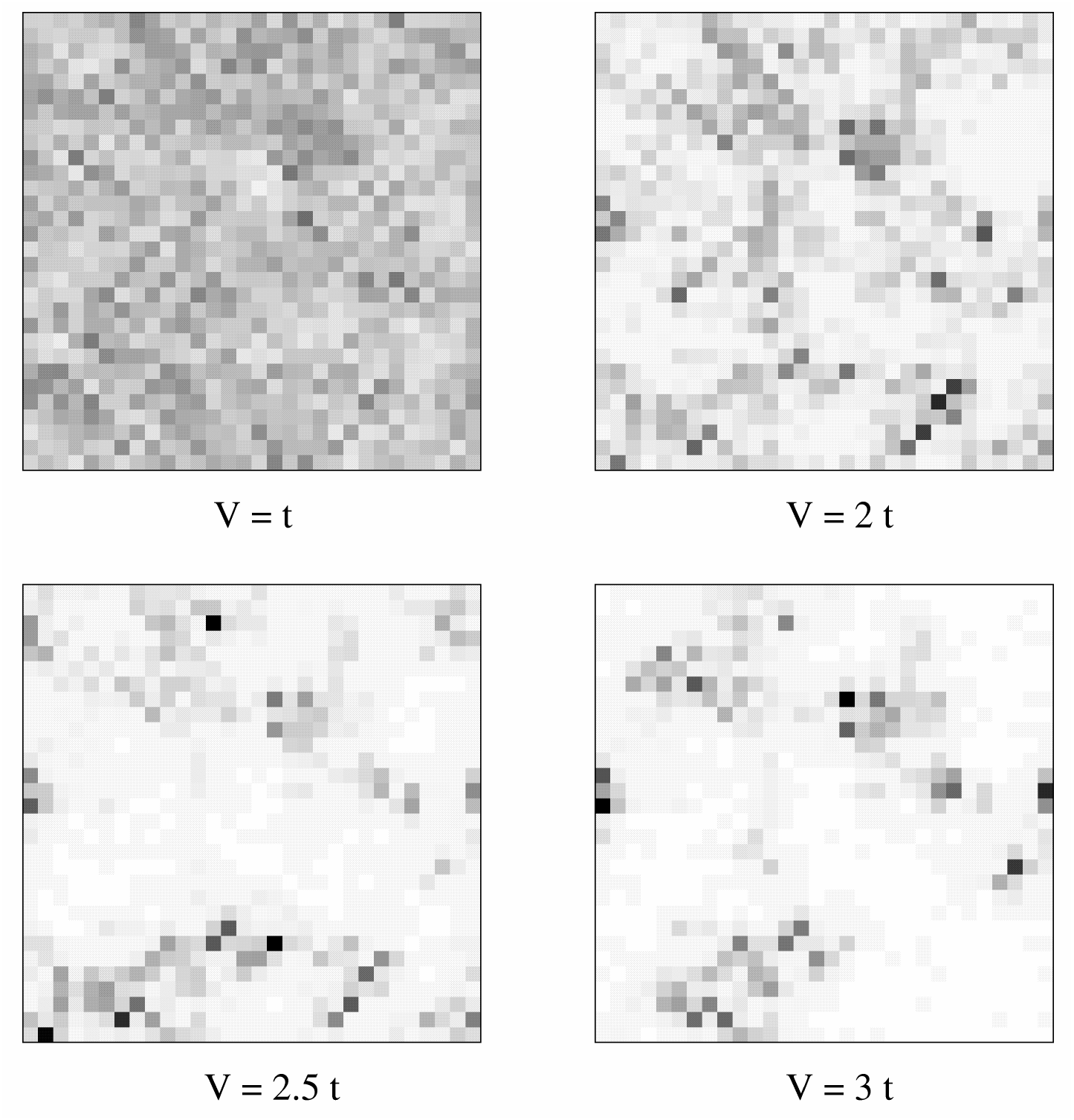}
	\caption{\label{ghosalblobs} Maps of the pairing amplitude $\Delta({\bf R}) = \langle c_{{\bf R}\downarrow}c_{{\bf R}\uparrow}\rangle$ for a given disorder realization with increasing disorder strength $V$, according to BdG calculations.
	For low disorder $V=1t$ the pairing amplitude is fairly uniform over the system. With increasing disorder the system breaks up into SC puddles with large $\Delta$ in an insulating sea with $\Delta\approx 0$.
	}
	\end{minipage}
	\hspace{2pc}
	\begin{minipage}{16pc}
	\includegraphics[width=16pc]{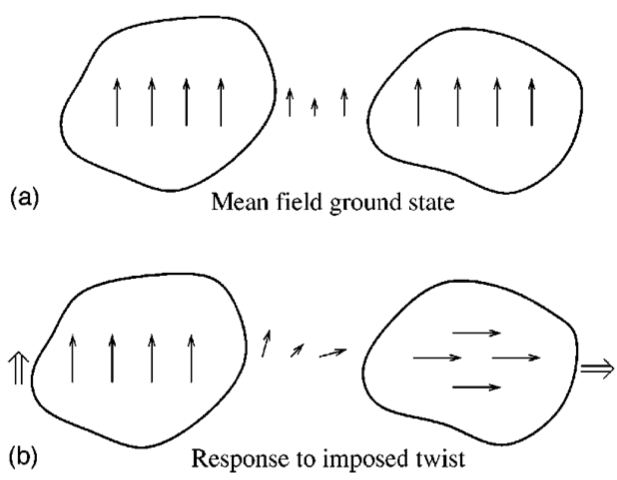}\hspace{2pc}%
	\caption{\label{GhosalPuddleCartoon} The superfluid density is a measure of the rigidity of the phases. This rigidity is clearly reduced by thermal fluctuations. It is also reduced by amplitude 
	variations and by quantum phase fluctuations even at $T=0$. The upper panel shows a phase coherent ground state albeit with variations in the amplitude-- large values in the SC puddles and small values in the intervening sea.
	For an applied twist, the system can accommodate most of the twist in regions where the amplitude is small leading to a very small cost in energy and hence a very small superfluid density. }
	\end{minipage}
\end{figure}

\begin{figure}[h]
	\centering
	\begin{minipage}{35pc}
	\centering
	\includegraphics[width=30pc]{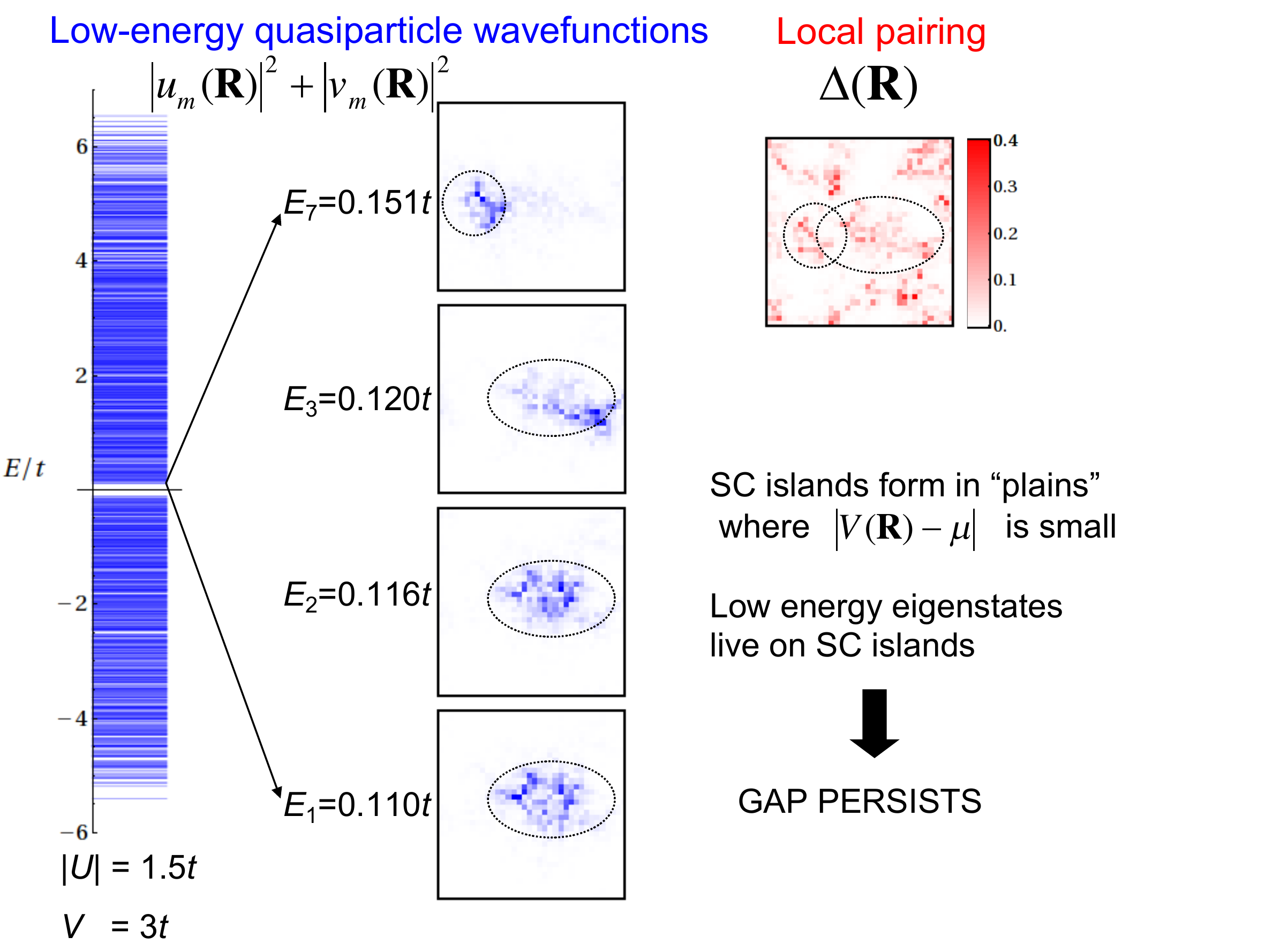}
	\caption{\label{eigenmodes} Eigenvalue spectrum for disorder $V=3t$. In the absence of any attraction, the spectrum is gapless at the chemical potential.
	However, even the smallest attraction ($U=-1.5t$) opens up a gap. In fact, as discussed in this paper, the gap remains finite for all values of the disorder.
	The eigenfunctions corresponding to the eigenvalues just above the gap edge correlate extremely well with SC puddles with relatively large pairing amplitude.
	For a given profile of random potentials, the system chooses those regions where $|V(r)-\mu|\approx 0$ to form the SC puddles.  On the other
	hand, eigenfunctions corresponding to the eigenvalues far below or above the gap edge correlate with deep valleys with a trapped pair or high mountains that are empty. These regions have a pairing amplitude close to zero due
	to strong localization effects.}
	\end{minipage}
\end{figure}

The key reason for the formation of a pseudogap regime even in a weakly coupled BCS superconductor is the phenomenon of disorder-induced ``emergent granularity'': a ``homogeneously disordered'' model or sample may have a strongly inhomogeneous pairing amplitude.  This emerged from Bogoliubov-de Gennes (BdG) calculations \cite{ghosal1998,ghosal2001} and is supported by QMC work and experiments.  See Figs.~\ref{ghosalblobs},\ref{eigenmodes},\ref{LocalSpecAndLocalGap}, and \ref{SacepeGapMap}.
 At weak disorder the pairing amplitude is homogeneous.  At strong disorder it breaks up into blobs.
The phases of the order parameter on different blobs are weakly coupled, so the system is similar to a granular superconductor or a Josephson Junction Array (JJA), where phase fluctuations are extremely important (see Fig.~\ref{GhosalPuddleCartoon}).  A treatment of phase fluctuations requires going beyond BdG.

\section{Thermal Fluctuations versus Quantum Fluctuations}
Thermal phase fluctuations include Goldstone modes and vortices.  These may be treated using Monte Carlo simulation of a XY or Ginzburg-Landau action for a Hubbard-Stratonovich auxiliary field~\cite{dubi2007,dubi2008}.

At $T=0$ thermal fluctuations are quenched.  However, long-range phase coherence can nevertheless be destroyed by quantum phase fluctuations. These can be understood to arise primarily as a consequence of the number-phase uncertainty.  In a bulk superconductor, the number of fermions is completely uncertain and phase is uniform.  Disorder confines the wavefunctions and places constraints on fermion number, thus introducing phase fluctuations that can ultimately drive the system into an insulator.
These fluctuations can be captured qualitatively with the self-consistent harmonic approximation~\cite{ghosal1998,ghosal2001, tvr,stroud} or more quantitatively using quantum Monte Carlo and maximum entropy methods.

\section{Minimal Model for the SIT}
To model the competition between superconductivity and localization that leads to the SIT in quench-condensed films with thicknesses less than the 
coherence length, we take the simplest lattice Hamiltonian that has an $s$-wave superconducting ground state in the absence of disorder ($V=0$)
and exhibits Anderson localization when the attractive interaction is turned off ($U=0$).
Thus, we study the two-dimensional attractive Hubbard model in a random potential:
	\begin{align}
	H&= -t\sum_{\langle \RRR \RRR' \rangle \sigma} 
			(\cdag_{\RRR\sigma} \cccc_{\RRR'\sigma} + \cdag_{\RRR'\sigma} \cccc_{\RRR\sigma})\nonumber \\
	&- \sum_{\RRR\sigma} (\mu-V_\RRR) n_{\RRR\sigma}
		-	|U|	\sum_\RRR n_{\RRR\uparrow} n_{\RRR\downarrow}.
	\label{Hamiltonian}
	\end{align}
with lattice sites $\RRR$ and $\RRR'$, spin indices $\sigma=\up$ or $\dn$, 
fermion creation and annihilation operators $\cdag_{\RRR\sigma}$ and $\cccc_{\RRR\sigma}$,  
number operators $n_{\RRR\sigma} = \cdag_{\RRR\sigma} \cccc_{\RRR\sigma}$,
hopping $t$ between neighboring sites $\langle \RRR\RRR' \rangle$,
and a chemical potential $\mu$ chosen such that the average density is $\mean{n} \neq 1$.
$V_\RRR$ is a random potential at each site drawn from the uniform distribution on $[-V,+V]$, 
and $\left| U \right|$ is the on-site attraction leading to $s$-wave SC.
We will measure all energies in units of $t$.

\section{Predictions from Theory}
Numerous studies taken together, including Bogoliubov-de Gennes (BdG) mean-field theory, determinant quantum Monte Carlo (QMC) simulations, and maximum entropy method (MEM) analytic continuation, give rise to the following predictions near the disorder-tuned SIT.

\subsection{Emergent Granularity}

There has been considerable debate about whether the transition is driven by a destruction of the pairing amplitude or the phase stiffness.
It was believed that homogeneously disordered films would show a closing of the single particle gap in tunneling signaling a destruction of the pairing amplitude,
whereas intentionally grown granular films would lose global superconductivity because of a loss of phase coherence between grains even though the individual grains could be locally superconducting. Our results indicate that, at least for this minimal model, there is a {\em single paradigm} for homogeneously disordered and granular films. 
For low disorder the local pairing amplitude $\Delta({\bf R})\equiv \langle c_{{\bf R}\downarrow}c_{{\bf R}\uparrow}\rangle$ is homogeneous across the system.
However, as disorder is increased the system self-organizes into superconducting blobs on the scale of the coherence length within an insulating matrix
(see Figs.~\ref{ghosalblobs} and~\ref{eigenmodes}). The phases of the different blobs are coupled by Josephson tunneling of pairs.
In the globally superconducting state, the phases of the different blobs get locked together whereas in the insulator the phase coherence of the different blobs is lost on ever shorter length and time scales as one moves away from the quantum phase transition.

\begin{figure}[htb]
\centering
\begin{minipage}{20pc}
\includegraphics[width=20pc,clip,trim=0 6pc 0 6pc]{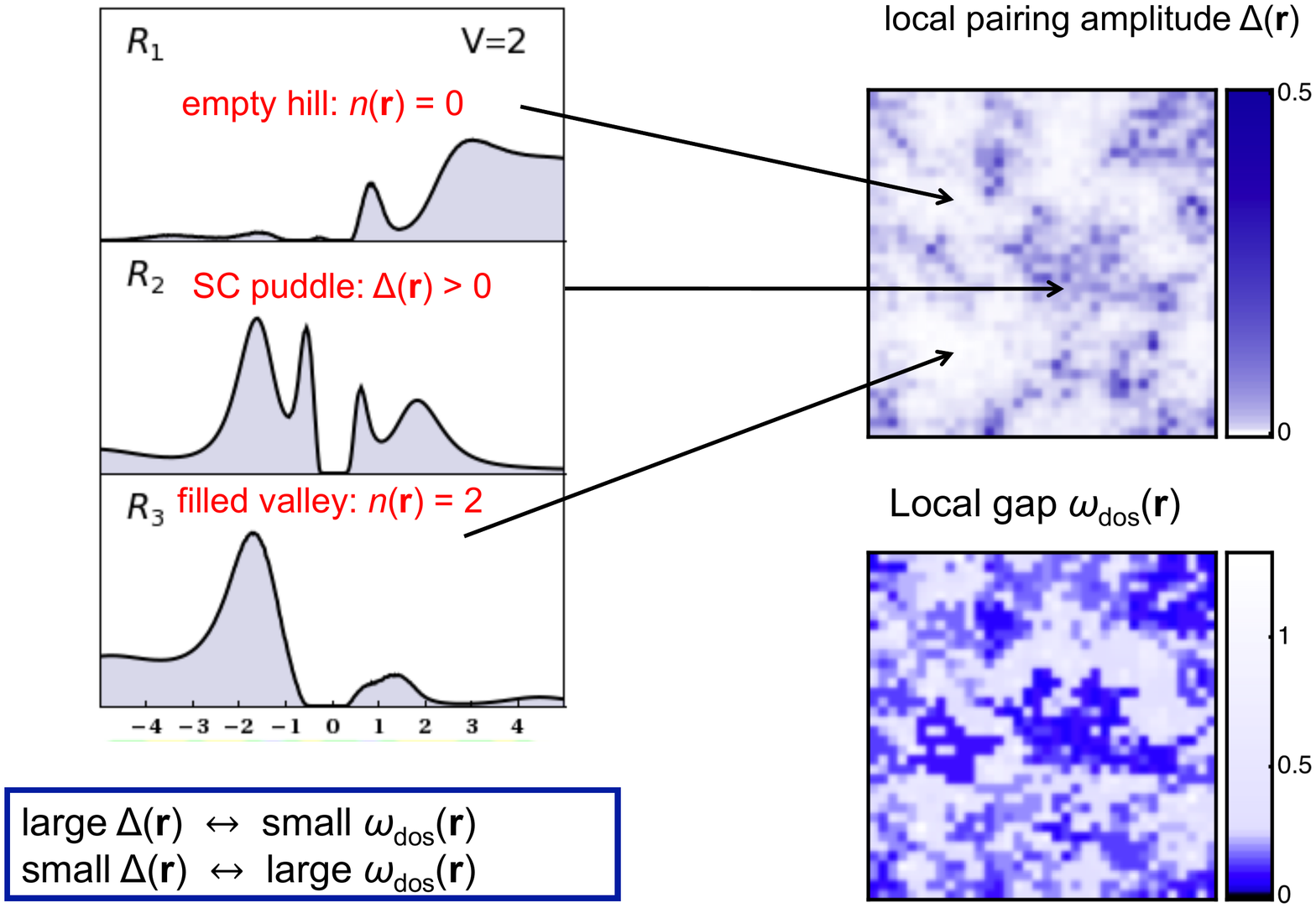}
\caption{\label{LocalSpecAndLocalGap} 
	(Right) Within BdG, the local pairing amplitude is \emph{anticorrelated} with the local gap.
	(Left) LDOS results from QMC+MEM.
	Site ${\bf R_1}$ is on a high potential hill that is nearly empty, 
	and ${\bf R_3}$ is in a deep valley that is almost doubly occupied. 
	This leads to the characteristic asymmetries 
	in the LDOS for ${\bf R_1}$ and ${\bf R_3}$. The small
	local pairing amplitude $\Delta(\RRR)$ at these two sites is reflected in the absence of coherence peaks in the LDOS.
	In contrast, site ${\bf R_2}$ has a density closer to half-filling,
	leading to a significant local pairing amplitude, a much more symmetrical LDOS, 
	and coherence peaks that persist even at strong disorder.
}
\end{minipage}
\hspace{1pc}
\begin{minipage}{14pc}
\includegraphics[width=14pc]{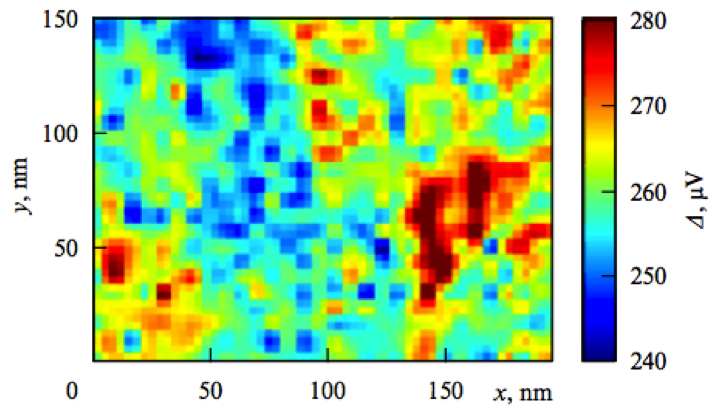}
\caption{\label{SacepeGapMap} Gap map of a TiN film obtained from scanning tunneling spectroscopy, showing inhomogeneities on a scale of a few tens of nanometers.}
\end{minipage}
\end{figure}

\subsection{Pairing amplitude does not equal gap}
In a clean system, the pairing amplitude equals the single-particle gap, and it is common to use the symbol $\Delta$ for both quantities.

In a disordered system near the SIT, the pairing amplitude $\Delta(\RRR)$ is strongly inhomogeneous.  Does it describe the local gap $E_\text{gap} (\RRR)$ as measured by scanning tunneling spectroscopy?  The answer is no!

Figure~\ref{LocalSpecAndLocalGap} shows that there is in fact an {anti-correlation} between the local pairing amplitude and the local spectral gap. The SC puddles on which the local $\Delta(\RRR)$ is finite have a finite 
gap with symmetric line shapes and sharp coherence peaks or pile-ups in the density of states at the gap edges. On the other hand, the insulating regions have $\Delta(\RRR)\approx 0$ and very asymmetric broad density of states showing a much larger gap. Although the local gap extracted from the local density of states (LDOS) 
is highly inhomogeneous, it is nevertheless finite at every site,
similar to the experimental data in Fig.~\ref{SacepeGapMap}.

\begin{figure}[h!]  \centering
\begin{minipage}[b]{24pc}
\includegraphics[width=13pc]{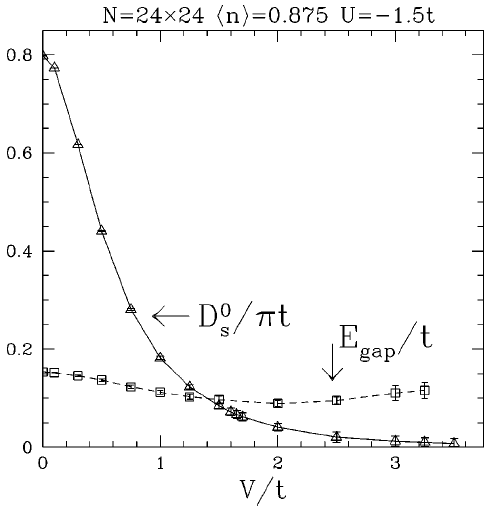}
\hspace{1pc}%
\includegraphics[width=7pc]{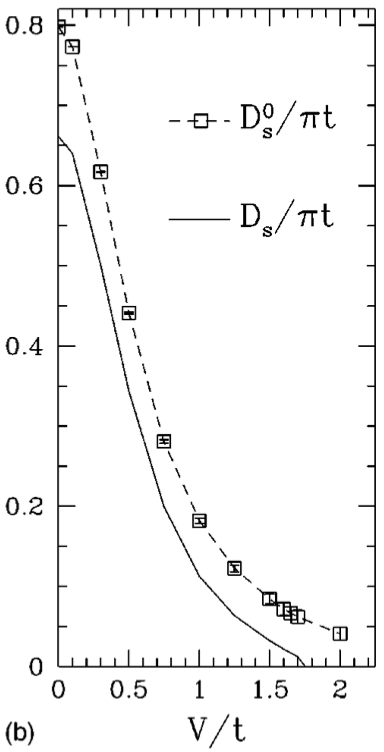}
\hspace{1pc}%
\end{minipage}
\begin{minipage}[b]{12pc}
\caption{\label{GhosalGapAndStiffness} 
(Left) According to BdG simulations, the gap persists across the SIT.  This conclusion holds even after phase fluctuations are taken into account.
(Right) The renormalized superfluid stiffness $D_s$, including phase fluctuations within the SCHA, falls to zero at the SIT.
}
\end{minipage}
\end{figure}

The DOS is the LDOS averaged over all sites.  The gap in the DOS, $E_{gap}$, is the lowest gap in the LDOS on any site.
According to BdG (Fig.~\ref{GhosalGapAndStiffness}) and QMC (Fig.~\ref{KarimPhaseDiagram}) calculations, $E_{gap}$ remains robust across the SIT, even when thermal and quantum phase fluctuations are included.  Thus the SIT is a transition from a gapped superconductor to a gapped insulator.

\begin{figure}[h]
	\centering
	\begin{minipage}{16pc}
		\includegraphics[width=16pc]{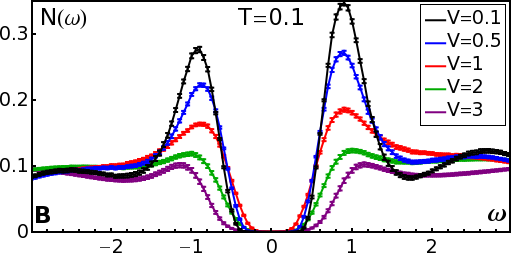}
	\end{minipage}
	\hspace{1pc}
	\begin{minipage}{16pc}
		\caption{
		\label{DisorderDependenceQMCMEM} 
		Disorder dependence of single-particle spectrum from QMC+MEM
		at very low temperature  \cite{bouadim2011}.
		With increasing disorder, 
		quantum phase fluctuations eventually wash out the coherence peaks,
		but the gap is robust.
		}
	\end{minipage}
\end{figure}

\subsection{Coherence peaks disappear beyond SIT}

These characteristic pile-ups in the DOS at the gap edges 
appear to be directly correlated with superconducting order.
They vanish as the disorder is increased across the SIT (Fig.~\ref{DisorderDependenceQMCMEM}),
or as the temperature is raised above $T_c$ (Fig.~\ref{TemperatureDependenceQMCMEM}).
These predictions agree very well with experiments (see Fig.~\ref{TemperatureDependenceInO}).

\subsection{Pseudogap over wide temperature range}

\begin{figure}[h]
	\centering
	\begin{minipage}{16pc}
	\includegraphics[width=16pc]{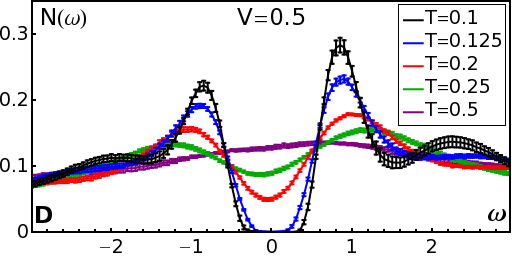}
	\includegraphics[width=16pc]{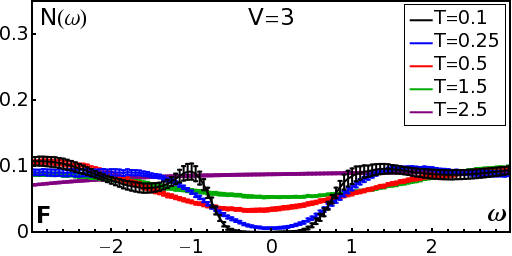}
	\caption{
	\label{TemperatureDependenceQMCMEM} 
	Temperature dependence of DOS from QMC+MEM calculations \cite{bouadim2011}.
	(Top) At weak disorder, as a function of increasing temperature,
	thermal fluctuations destroy the coherence peaks for $T \gtrsim T_c \approx 0.14$.
	However, a pseudogap remains up to higher temperatures $T \sim 0.4$.
	(Bottom) At strong disorder, there are no coherence peaks;
	there is a hard gap at $T=0$ and a pseudogap up to $T \sim 1.5$.
	}
	\end{minipage}
	\hspace{2pc}
	%
	\begin{minipage}{16pc}
	\includegraphics[width=16pc]{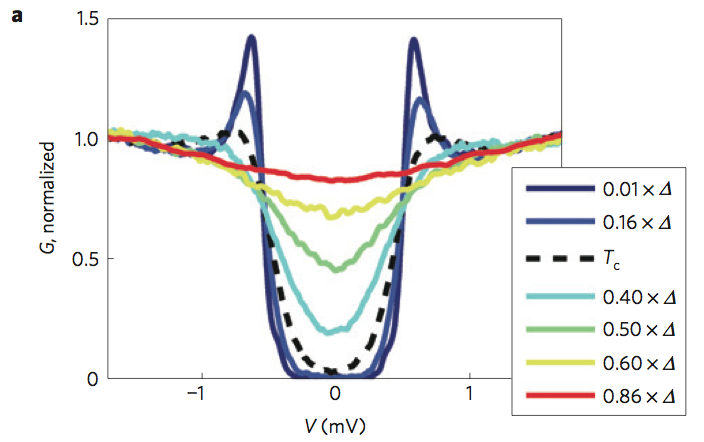}
	\includegraphics[width=13pc]{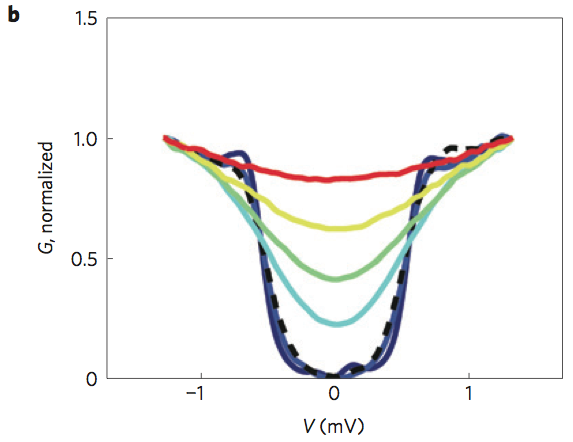}
	\caption{
	\label{TemperatureDependenceInO} 
		Local tunneling conductance spectrum in InO films	\cite{sacepe2011}.
		(Top) At low disorder	 there are coherence peaks below $T_c$
		 and a pseudogap up to much higher $T$.		
		(Bottom) At high disorder there are no coherence peaks, 
			but nevertheless there is a hard gap at low $T$
			and a pseudogap up to high temperatures.
	}
	\end{minipage}
\end{figure}

Near the SIT, a pseudogap -- a suppression in the low-energy DOS -- 
persists well above the superconducting $T_c$ up 
to a crossover temperature scale $T^\ast$, in marked deviation from BCS theory.
This disorder-driven pseudogap also exists at finite temperatures in the insulating state
and grows with disorder (Fig.~\ref{TemperatureDependenceQMCMEM}.)
These predictions are again in good agreement with experiments 
(Fig.~\ref{TemperatureDependenceInO}).

\begin{figure}[h]
\centering
\begin{minipage}{16pc}
\includegraphics[width=13pc]{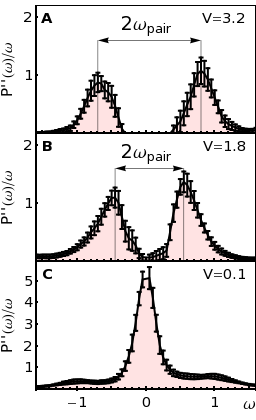}
\caption{\label{KarimPairSuscep}
	Imaginary part of the dynamical pair susceptibility $P''(\omega) / \omega$ at $T=0.1t$.
	Error bars represent variations between 10 disorder realizations.
	For $V<V_c$ the large peak at $\omega=0$
	 indicates zero energy cost to insert a pair into the SC.
	For $V>V_c$, there is a gap-like structure at $\pm \omega_{\rm pair}$,
	the typical energy required to insert a pair into the insulator.
}
\end{minipage}
\hspace{2pc}
%
\begin{minipage}{16pc}
\includegraphics[width=14pc]{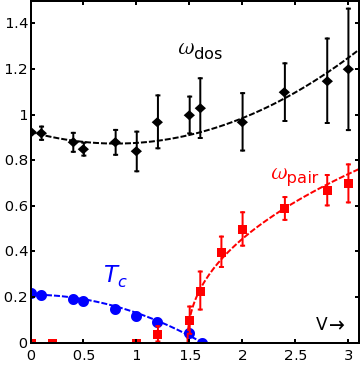}
\caption{\label{KarimPhaseDiagram}
	Energy and temperature scales across the superconductor-insulator transition (SIT)
	according to QMC+MEM calculations.
	The single-particle gap $\omegados$ remains finite for all values of disorder $V$,
	whereas the superconducting $T_c$ 
	and the two-particle energy scale $\omega_{\rm pair}$ in the insulator
	both vanish at the SIT.
}
\end{minipage} 
\end{figure}

\subsection{Two-particle spectrum}
The local two-particle spectral function, or pair susceptibility $P(\omega)$, is defined as the analytic continuation of the correlation function $P(\tau) = \sum_\RRR \mean{\calT_\tau F (\RRR;\tau) F^\dag (\RRR;0)}$ where $F(\RRR,\tau)=  \cccc_{\RRR\dn} (\tau) \cccc_{\RRR\up} (\tau)$.  Physically, $P(\RRR,\omega)$ is the amplitude for inserting a pair at a site $\RRR$ at energy $\omega$, and $P(\omega)$ is the average insertion amplitude over all sites.


Fig.~\ref{KarimPairSuscep} shows QMC+MEM results for the imaginary part of $P$.
On the superconducting side of the transition there is a large amplitude for inserting pairs at zero energy.
However, on the insulating side, there is a characteristic energy scale $\omega_{\rm pair}$ to insert a pair in the insulator that collapses upon approaching the SIT
(notwithstanding a small amount of spectral weight at low energies coming from rare regions).

\subsection{Suggestions for Further Experiments}
A large number of experiments exploring SITs in conventional $s$-wave materials have focused on transport measurements.  In order to get deeper insights into the phases, 
it is useful to bring to bear a larger variety of experiments of SIT, such as a detailed and systematic analysis of scanning tunneling spectroscopy and dynamical conductivity, some of which have already begun. In the cuprates, considerable insight about the pseudogap region was obtained by measuring
the entropy carried by vortices from the Nernst effect~\cite{wangOng2003}.  Similar experiments for the regular SIT should also be extremely useful~\cite{pourret2006}.
It is evident that there should be large diamagnetic effects in the insulator because of puddles of superconducting regions embedded in an insulating matrix.
Once again direct probes of diamagnetic in the insulator and its evolution toward the SIT, similar to the explorations in the cuprates\cite{ongDiamagnetism2005,ongDiamagnetism2010}, should be extremely instructive. Unlike the disorder driven transition, the effect of a perpendicular magnetic field is not as well understood. The experiments show that the sharp normal to SC transition at $T_c$ for zero field develops long tails in the presence of a magnetic field. In other words there appears to be some source of dissipation in the SC state. Is this due to unpinned vortices? Why is the disorder, even when it is as high as the quantum of resistance, unable to pin the vortices? 

And finally, once the phases are understood, the next frontier will be investigations of the quantum phase transition. In this context the ability to tune across the transition by gating is a breakthrough!
Gating allows one to separately tune the effects of screening and disorder and will hopefully provide a much better understanding of quantum criticality.

\subsection{Acknowledgments}
We gratefully acknowledge support from  US Department of Energy, 
Office of Basic Energy Sciences grant DOE DE-FG02-07ER46423 (N.T., Y.L.L.), 
NSF DMR-0907275 (K.B.), NSF DMR-0706203 and NSF DMR-1006532 (M.R.), and computational support from 
the Ohio Supercomputing Center.

\section*{References}


\providecommand{\newblock}{}

\end{document}